\documentclass{jfm}

\usepackage{graphicx}
\usepackage{epstopdf,epsfig}
\usepackage{newtxtext}
\usepackage{newtxmath}
\usepackage{natbib}
\usepackage{hyperref}
\usepackage{siunitx}
\usepackage{amssymb,bm}
\usepackage{xcolor}
\usepackage{caption}

\usepackage{float}
\hypersetup{
    colorlinks = true,
    urlcolor   = blue,
    citecolor  = black,
}

\newcommand{\RomanNumeralCaps}[1]

\shorttitle{Trapping microswimmers in acoustic streaming flow}
\shortauthor{X. Sun, W. Tan and Y. Man}

\title{Trapping microswimmers in acoustic streaming flow}

\author{Xuyang Sun\aff{1}, Wenchang  Tan\aff{1}\aff{2}\aff{3}\aff{4}\corresp{\email{tanwch@pku.edu.cn}} \and Yi Man\aff{1}\aff{2}\corresp{\email{yiman@pku.edu.cn}}}

\affiliation{\aff{1}Department of Mechanics and Engineering Science, College of Engineering, Peking
University, Beijing, China
\aff{2}State Key Laboratory for Turbulence and Complex Systems, Peking University, Beijing, China
\aff{3}PKU-HKUST Shenzhen-Hong Kong Institution, Shenzhen, Guangdong, China
\aff{4}Shenzhen Graduate School, Peking University, Shenzhen, Guangdong, China}

\begin{document}
\maketitle

\begin{abstract}
The acoustofluidic method holds great promise for manipulating microorganisms. When exposed to the steady vortex structures of acoustic streaming flow, these microorganisms exhibit intriguing dynamic behaviors, such as hydrodynamic trapping and aggregation. To uncover the mechanisms behind these behaviors, we investigate the swimming dynamics of both passive and active particles within a two-dimensional acoustic streaming flow. By employing a theoretically calculated streaming flow field, we demonstrate the existence of stable bounded orbits for particles. Additionally, we introduce rotational diffusion and examine the distribution of particles under varying flow strengths. Our findings reveal that active particles can laterally migrate across streamlines and become trapped in stable bounded orbits closer to the vortex center, whereas passive particles are confined to movement along the streamlines. We emphasize the influence of the flow field on the distribution and trapping of active particles, identifying a flow configuration that maximizes their aggregation. These insights contribute to the manipulation of microswimmers and the development of innovative biological microfluidic chips.

\end{abstract}

\section{Introduction}

Microswimmers in flow exhibit a wide range of fascinating behaviors, such as directional alignment \citep{Hope2016}, hydrodynamic trapping \citep{Sipos2015} and aggregation \citep{Torney2007}, all of which are heavily influenced by their surrounding flow environment. The hydrodynamic interactions between microswimmers' activity and the flow field play a crucial role in shaping these dynamics \citep{eric2016,math2019,baker2019}. Understanding these hydrodynamic interactions is critical, as they not only govern microswimmer behavior but also have important applications in areas like targeted drug delivery \citep{Park2017}, bioengineering \citep{Liu2022}, and environmental monitoring \citep{Liang2018}. As a result, investigating the physical mechanisms behind microswimmer-flow interactions, and leveraging this knowledge for microswimmer manipulation, has become a key focus in the study of active matter \citep{bech2016,whee2019,aran2022}.

Early research on swimming behaviors of microswimmers primarily focused on planar shear flow environments \citep{zottl2012,Kantsler2014,lee2021,Rubio2021}.
When bacteria move in the shear flow, they tend to accumulate in regions of high shear rate \citep{Rusconi2014}, which also is theoretically demonstrated by \citet{Vennamneni2020}. They find that the interaction between the shear rate and bacteria is the crucial factor for achieving migratory motion and aggregation behavior. More recently, the study of microswimmers has been extended to their swimming dynamics in vortex flows\citep{simon2020}. For instance, \citet{marcos2006} demonstrate that bacteria align with streamlines in strong vortices, a result consistent with our findings for passive particles. For gyrotactic microorganisms, such as algae, studies by Durham et al. \citep{durham2011,durham2013} have highlighted how gyrotaxis influences their aggregation dynamics in vortical flows. Furthermore, \citet{Sokolov2016} find that elongated bacteria in the vortex flow generated by a rotating microparticle cross the streaminline and form a depletion zone near the vortex core. \citet{ivan2022} have employed a theoretical model to demonstrate that the depletion zone is caused by the interaction between particles and the vortex flow. However, in the Gaussian vortex flow field, the trapped active particles aggregate around the vortex core to form high-density region \citep{jos2020}. Additionally, dumbbell-shaped particles exhibit spiral motion in a two-dimensional steady vortex flow, with the shape of their orbits depending on their initial position and direction \citep{Yerasi2022}.

           \begin{figure}
           \centerline{
            \includegraphics[width=8cm]{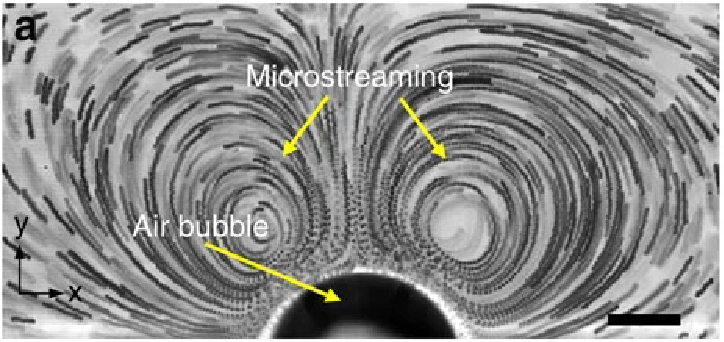}}
            \captionsetup{justification=justified, singlelinecheck=false}
            \caption{The optical image of the acoustic streaming flow generated by the oscillating microbubble in experiment, reproduced from \citet{ahmed2016} under terms of the CC-BY 4.0. licence. Copyright 2016 The Authors, published by Springer Nature.}
             \label{fig1}
           \end{figure}

Given the interaction between the particles and background flow field, the use of microfluidic chips has attracted widespread interest in the biomedical field \citep{Collins2017, lu2020, Geng2023, Hazal2024, Liu2024}. The acoustic streaming flow generated by ultrasonically driven an trapped microbubble is a primary method used in microfluidic chips for manipulating particles \citep{Marmottant2004,Chen2016,zhang2020}.
As shown in the figure \ref{fig1}, the trapped cavitated microbubble undergoes radial and tangential oscillations under ultrasonic excitation, producing a stable acoustic streaming flow consisting of a pair of vortices \citep{ahmed2016}. \citet{Rogers2011} find that the behavior of passive particles within the acoustic streaming flow are influenced by their size and density, making this method widely applicable in the separation of cells \citep{Li2021,Gao2022}. Additionally, experimental evidence demonstrates that \textit{Escherichia coli} can be collected near the vortex center of the acoustic streaming flow generated by an oscillating bubble, leading to the formation of a biofilm structure \citep{yazdi2012}. Further experiment has demonstrated that multiple oscillating microbubbles of varying sizes can create an acoustic network flow, which facilitates the non-destructive pumping of \textit{Escherichia coli} \citep{gao2020}. Despite these advancements, the effects of acoustic streaming flow on the swimming dynamics and aggregation behaviors of active particles, as well as the underlying physical mechanisms governing these processes, remain poorly understood and require further elucidation.

Inspired by the observed aggregation behavior of bacteria in an acoustic streaming field during experiments \citep{yazdi2012}, we have conducted a detailed investigation into the swimming dynamics of active particles in the acoustic streaming flow generated by an oscillating microbubble. We employ an approximate solution for the acoustic streaming flow \citep{spelman2015}, and then we establish a deterministic model to predict the trapped bounded orbits of both passive and active particles. Our results shows that active particles are trapped in the bounded orbits near the vortex center due to their activity, while passive particles are restricted to moving along the streamlines. The stable motion region of active particle depends on the streaming flow, which further proves the dependence of active microorganisms on the environment. To further understand the behavior of particle in more realistic environment, we introduce rotational noise to analyze the trapping and distribution of both passive and active particles. We find that active particles exhibit a non-uniform distribution within the streaming flow and we identify a flow configuration that maximizes the degree of aggregation of the active particles. Our results provide insights for the use of microfluidic chips in rapid detection and separation of microorganisms.

\section{Setup and Acoustic streaming flows}

          \begin{figure} \centerline{\includegraphics[width=13.5cm]{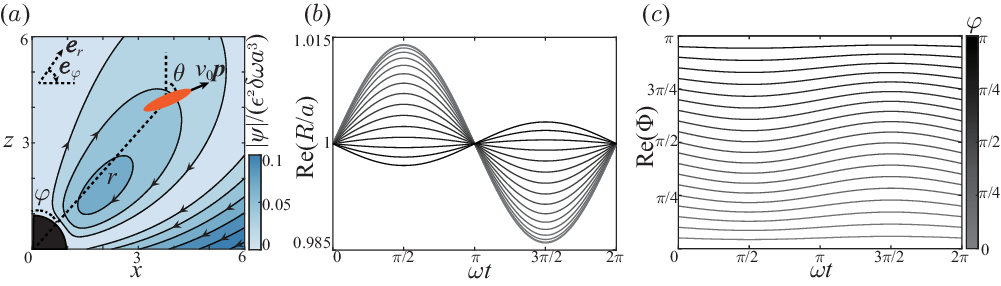}}
            \captionsetup{justification=justified, singlelinecheck=false}
            \caption{($a$) Illustration of a microswimmer moves in the acoustic streaming flow generated by an oscillating bubble. The microswimmer is depicted as an orange ellipsoid particle located at ($r$,$\varphi$) with constant velocity $v_0$ along $\bm{p}$. The magnitude of scaled streaming flow is $|\psi|/(\epsilon^2\delta\omega a^3)$ at $V_0=0.4$, $V_1=1$. ($b$)-($c$) The time evolution results of the real part of boundary position $R$ and $\Phi$ of the oscillating bubble for one period based on the same parameters with representative value $\epsilon=0.01$.}
             \label{fig2}
             \end{figure}
\subsection{Setup and nondimensionalization}
The microswimmer is modeled as a spheroidal particle characterized by an aspect ratio $\lambda$, defined as the ratio of the long axis to the short axis. We assume that the swimmer maintains a constant intrinsic swimming speed $v_0\bm{p}$, where $\bm{p}$ denotes the swimmer's orientation. For an elongated swimmer, the orientation aligns with the direction of the long axis.

The streaming flow is generated by an acoustically-driven oscillating microbubble with radius $a$. For simplicity, we assume that the bubble is fixed in an unbounded domain and ignore wall effects. In addition, we assume that the motion of the bubble's surface is axisymmetric, incorporating both radial and tangential oscillations. The surface of the microbubble is presumed to oscillate at an angular frequency $\omega$ with a small amplitude $\epsilon a$, where $\epsilon \ll 1$. We employ spherical coordinates $(r,\varphi)$ as illustrated in figure \ref{fig2}. Utilizing a Lagrangian framework, the boundary shape of the microbubble is described by its radial position $R$ and angular position $\Phi$, measured at the location of a material point initially at $(a,\varphi)$. Due to the axisymmetry, the surface oscillation modes can be expressed as a series of orthogonal basis functions formed by Legendre polynomials, presented as follows:
 \begin{align}\label{eq01}
         \aligned
           &R(\varphi,t)=a-\epsilon a\sum^{\infty}_{n=0}V_nP_n(\cos\varphi)e^{i(\omega t+\frac{\pi}{2})},\\
           &\Phi(\varphi,t)=\varphi+\epsilon\sum^{\infty}_{n=1}W_n\left(\frac{\int_{\cos\varphi}^1 P_n(x)dx}{\sin\varphi}\right)e^{i(\omega t+\frac{\pi}{2})},
         \endaligned
 \end{align}
where $V_n$ and $W_n$ are arbitrary complex constants determined by the surface motion,  and $P_n(x)$ is the Legendre Polynomial of degree $n$. The flow field is governed by the Navier-Stokes equation:
        \begin{align}\label{eq02}
         \aligned
            &\frac{\partial \bm{u}}{\partial t}+(\bm{u}\cdot\bm{\nabla})\bm{u}=-\frac{1}{\rho}\bm{\nabla} p+\nu\bm{\nabla}^2\bm{u},
         \endaligned
        \end{align}
where $\bm{u}=(u_r, u_{\varphi})$ represents the fluid velocity, $p$ is the pressure. Since the flow generated by the oscillating bubble is axisymmetric, we introduce a stream function $\psi(r,\varphi)$. In spherical coordinates, the fluid velocities are represented as follows:
        \begin{align}\label{eq03}
         \aligned
           &u_r=\frac{1}{r^2\sin\varphi}\frac{\partial \psi}{\partial \varphi},\quad u_{\varphi}=-\frac{1}{r\sin\varphi}\frac{\partial \psi}{\partial r}.
         \endaligned
        \end{align}
Substituting equation (\ref{eq03}) into (\ref{eq02}), Navier-Stokes equation becomes
        \begin{align}\label{eq04}
         \aligned
           &\frac{\partial \left(D^2\psi\right)}{\partial t}+\frac{1}{r^2}\left[-\frac{1}{\sin\varphi}\frac{\partial \psi}{\partial r}\frac{\partial \left(D^2\psi\right)}{\partial \varphi}+\frac{1}{\sin\varphi}\frac{\partial \psi}{\partial \varphi}\frac{\partial \left(D^2\psi\right)}{\partial r}+2D^2\psi L\psi\right]=\nu D^2\left(D^2\psi\right),
         \endaligned
        \end{align}
where the operators are
        \begin{align}\label{eq05}
         \aligned
           &D^2=\frac{\partial^2}{\partial r^2}+\frac{1}{r^2}\left(\frac{\partial^2}{\partial {\varphi}^2}-\frac{\cos\varphi}{\sin\varphi}\frac{\partial}{\partial \varphi}\right),\quad L=\frac{\cos\varphi}{\sin^2\varphi}\frac{\partial }{\partial r}-\frac{1}{r\sin\varphi}\frac{\partial}{\partial \varphi}\cdot
         \endaligned
        \end{align}
On the surface of the bubble, the fluid velocity satisfies the boundary condition:
       \begin{align}\label{eq06}
          \aligned
           &u_{r}(R, \Phi)=\frac{\partial R}{\partial t}\bigg|_{(R, \Phi)}, \quad u_{\varphi}(R, \Phi)=R\frac{\partial \Phi}{\partial t}\bigg|_{(R, \Phi)}.
         \endaligned
        \end{align}
In addition, the incorporation of an additional boundary condition is crucial in order to guarantee the absence of any tangential stress exerted on the surface of an oscillating bubble:
        \begin{align}\label{eq07}
         \aligned
           &\left[\frac{1}{r}\frac{\partial u_r}{\partial \varphi}+\frac{\partial u_{\varphi}}{\partial r}-\frac{u_{\varphi}}{r}\right]\bigg|_{(R, \Phi)}=0.
         \endaligned
        \end{align}
We will sketch the asymptotic solution of \eqref{eq04} in Section \ref{sec:flow}. We assume that the disturbance of the flow field due to the active particle is negligible. Additionally, the hydrodynamic interactions between the active particle and the bubble surface are neglected in this study, as their effects are significantly weaker compared to the dominant influence of the acoustic streaming flow \citep{Spagnolie2015}. Consequently, the motion of the particle, as it drifts with the flow, and the rotation of the particle can be determined by Jeffery's equation \citep{jeffery1922}. The position of the particle $\bm{r}$ and the orientation $\bm{p}$ are governed by the following equations:
\begin{align}\label{eq08}
        \aligned
            &\frac{d \bm{r}}{\partial t}= v_0\bm{p}+\bm{u},\\
             &\frac{d \bm{p}}{\partial t}=\frac{1}{2}(\bm{\nabla} \times \bm{u})\times \bm{p}+\gamma(\bm{E}\cdot\bm{p}-(\bm{p}\cdot\bm{E}\cdot\bm{p})\bm{p}),
        \endaligned
\end{align}
where $\bm{E}=\left(\bm{\nabla}\bm{u}+\bm{\nabla}\bm{u^T}\right)/2$ is the strain rate tensor of the flow field, $\gamma=\left(\lambda^2-1\right)/\left(\lambda^2+1\right)$ represents the shape parameter of particle, $\lambda$ represents the ratio of the long axis to the short axis of microswimmer, $(-1<\gamma<0)$, $(\gamma=0)$ and $(0<\gamma<1)$ represent oblate spheroid, sphere and prolate ellipsoid, respectively.

Here, we firstly focus on obtaining solutions of the acoustic streaming flow field, we choose $\omega^{-1}$ as the relevant time scale and the radius $a$ of the oscillating bubble as characteristic length to non-dimensionalize the equation \eqref{eq04}. Consequently, the dimension of the stream function is $\omega a^3$. The dimensionless form of the equation \eqref{eq04} is then as follows:
        \begin{align}\label{eq09}
         \aligned
           &\frac{\partial \left(D^2\psi\right)}{\partial t}+\frac{1}{r^2}\left[-\frac{1}{\sin\varphi}\frac{\partial \psi}{\partial r}\frac{\partial \left(D^2\psi\right)}{\partial \varphi}+\frac{1}{\sin\varphi}\frac{\partial \psi}{\partial \varphi}\frac{\partial \left(D^2\psi\right)}{\partial r}+2D^2\psi L\psi\right]=\delta^2 D^2\left(D^2\psi\right).
         \endaligned
        \end{align}
Here, $\delta^2 = \nu/\left(\omega a^2\right)$, where $\delta$ represents the ratio of the viscous penetration length scale, $\left(\nu/\omega\right)^{1/2}$, to the bubble radius, $a$. The kinematic viscosity of a Newtonian fluid is typically $10^{-6}\,\si{m^2/s}$. In experiments, the range of $\omega a$ is $O\left(10^5\right)\sim O\left(10^7\right) \si{\micro\meter/s}$ \citep{yazdi2012}. The bubble radius is typically set to be $a=10\,\si{\micro\meter}$. Therefore, we can estimate that the range of $\delta$ is $O\left(10^{-1}\right) \sim O(1)$, which means that the viscous penetration length is smaller compared to the radius of the oscillating bubble. Furthermore, the relative oscillating amplitude $\epsilon$, is experimentally measured to be $O\left(10^{-2}\right)\sim O\left(10^{-1}\right)$. Thus, we assume the asymptotic limit $\epsilon \ll \delta \ll 1$, which
 is used to outline the asymptotic analysis in the following section. These estimates of $\epsilon$ and $\delta$ are primarily provided to illustrate that the asymptotic conditions can be reasonably satisfied, ensuring that the analytical derivation aligns with plausible experimental parameter ranges. Detailed discussions can be found in \citet{spelman2015}.

\subsection{Asymptotic solution of the streaming flow}\label{sec:flow}
In the section, we outline the main steps to calculate the asymptotic solution of the acoustic streaming flow, reproducing the known results from \citet{spelman2015} to ensure consistency with foundational work and to provide necessary details for the subsequent analysis of particle dynamics. Given that the oscillation of the bubble is of the order of $\epsilon$, it is straightforward to expand the flow field in orders of $\epsilon$:
        \begin{align}\label{eq10}
         \aligned
           &\left\{\psi,u_r,u_{\varphi}\right\}=\epsilon \left\{\psi^{(1)}, u^{(1)}_r, {u}^{(1)}_{\varphi}\right\}+\epsilon^2\left\{\psi^{(2)}, u^{(2)}_r, {u}^{(2)}_{\varphi}\right\}+o\left(\epsilon^3\right).
         \endaligned
        \end{align}
At the order of $\epsilon$, \eqref{eq09} is expressed as follows:
        \begin{align}\label{eq11}
        \aligned
           &\frac{\partial \left(D^2\psi^{(1)}\right)}{\partial t}=\delta^2 D^4\psi^{(1)}.\\
         \endaligned
        \end{align}
For the boundary conditions, we consider three modes in \eqref{eq01}: $(V_0, V_1, W_1)$. These modes are minimal yet sufficient to generate a streaming flow \citep{longuet1998}. By expanding \eqref{eq06} and \eqref{eq07}, we obtain the boundary conditions at the first order:
        \begin{align}\label{eq12}
         \aligned
           &\frac{\partial \psi^{(1)}}{\partial \varphi}\bigg|_{r=1}=[V_0+V_1\cos\varphi]\sin\varphi e^{i t},\\
           &\frac{\partial \psi^{(1)}}{\partial r}\bigg|_{r=1}=-\frac{1}{2} W_1\sin^2\varphi e^{i t} ,\\
           &\left(-\frac{\partial^2 \psi^{(1)}}{\partial r^2}+2\frac{\partial \psi^{(1)}}{\partial r}+\frac{\partial^2\psi^{(1)}}{\partial {\varphi}^2}-\frac{\cos\varphi}{\sin\varphi}\frac{\partial \psi^{(1)}}{\partial \varphi}\right)\bigg|_{r=1}=0.
         \endaligned
        \end{align}

In view of the oscillatory boundary condition \eqref{eq12}, the first-order solution will be $\psi^{(1)} \propto e^{it}$. By applying separation of variables twice, we can solve for the first-order stream function:
       \begin{align}\label{eq13}
         \aligned
           &\psi^{(1)}=e^{i t}\left[V_0(1-\cos\varphi)+\frac{1}{2}\left(\frac{2A\delta^2\sqrt{r}}{(1+i)^2}K_{\frac{3}{2}}\left(\frac{(1+i)r}{\sqrt{2}\delta }\right)+B\frac{1}{r}\right)\sin^2\varphi\right]\\
           &A=\frac{-1}{K_{\frac{3}{2}}\left(\frac{1+i}{\sqrt{2}\delta }\right)}\left(\frac{1}{\sqrt{2}\delta}(1+i)+1+\sqrt{2}\delta\frac{(i-1)}{4}\right)(W_1+V_1)+O\left(\delta^2\right),\\
           &B=V_1+\frac{\sqrt{2}\delta(1-i)}{2}(W_1+V_1)-i\delta^2(W_1+V_1)+O\left(\delta^3\right),\\
           &W_1=-V_1+\sqrt{2}\delta\frac{6}{1+i}V_1+18 i\delta^2 V_1+O\left(\delta^3\right),
         \endaligned
        \end{align}
where $K_{\frac{3}{2}}(r)$ is the modified Bessel function of the second kind of order $3/2$. As we can see, the coefficient $W_1$ is determined by $V_1$ as a result of the zero tangential stress condition \eqref{eq07} imposed on the bubble surface. To obtain a non-zero time-averaged flow, we need to consider the next order.  At the second order, the equation becomes:
      \begin{align}\label{eq14}
        \aligned
            &-\frac{\partial \left(D^2\psi^{(2)}\right)}{\partial t}+\delta^2 D^2\left(D^2\psi^{(2)}\right)\\
            &=\frac{1}{r^2}\left[-\frac{1}{\sin\varphi}\frac{\partial \psi^{(1)}}{\partial r}\frac{\partial \left(D^2\psi^{(1)}\right)}{\partial \varphi}+\frac{1}{\sin\varphi}\frac{\partial \psi^{(1)}}{\partial \varphi}\frac{\partial \left(D^2\psi^{(1)}\right)}{\partial r}+2D^2\psi^{(1)} L\psi^{(1)}\right].
         \endaligned
        \end{align}
This equation cannot be solved explicitly. Instead, we consider its time-averaged solution, $\langle \psi^{(2)} \rangle$, which is independent of time and satisfies:
       \begin{align}\label{eq15}
         \aligned
           & \delta^2 D^4\left\langle\psi^{(2)}\right\rangle=\left\langle\frac{1}{r^2}\left[-\frac{1}{\sin\varphi}\frac{\partial \psi^{(1)}}{\partial r}\frac{\partial \left(D^2\psi^{(1)}\right)}{\partial \varphi}+\frac{1}{\sin\varphi}\frac{\partial \psi^{(1)}}{\partial \varphi}\frac{\partial \left(D^2\psi^{(1)}\right)}{\partial r}+2D^2\psi^{(1)} L\psi^{(1)}\right]\right\rangle.
         \endaligned
        \end{align}
\eqref{eq15} can be solved using matched asymptotics. Since the interaction between the bubble and the fluid occurs over a short range, we consider a boundary layer adjacent to the bubble, with a thickness on the order of $\delta$. The distance from the bubble is rescaled as $\eta = (r-1)/\delta$. When we scale \eqref{eq15}, the left-hand side is of the order of $1/\delta^2$, and the right-hand side can be carefully expanded in orders of $\delta$, with the leading term being of the order of $1/\delta$. Consequently, the governing equation in the inner region has the following structure:
\begin{align}\label{eq16}
         \aligned
    \frac{\partial^4\left\langle\psi^{(2)}\right\rangle}{\partial \eta^4}&=\left[\delta C_1+\delta^2 (C_{21}+C_{22}\eta)\right]e^{-\frac{1+i}{\sqrt{2}}\eta}\sin^2\varphi+\left[\delta D_1+\delta^2 (D_{21}+D_{22}\eta)\right]e^{-\frac{1+i}{\sqrt{2}}\eta}\\&\sin^2\varphi\cos\varphi
    +\delta^2E_2e^{-\frac{1-i}{\sqrt{2}}\eta}\sin^2\varphi\cos\varphi+O\left(\delta^3\right),
         \endaligned
        \end{align}
where $C_1, C_{21}, C_{22}, D_1, D_{21}, D_{22}, E_2$ are complex constants that are functions of $V_0$ and $V_1$. Correspondingly, we rescale and expand the boundary conditions. The no-slip boundary condition can then be written in the following form:
        \begin{align}\label{eq17}
         \aligned
           \frac{\partial \left\langle\psi^{(2)}\right\rangle}{\partial \varphi}\bigg|_{\eta = 0}&=\left(a_0+\delta a_1+\delta^2 a_2\right)\sin\varphi\cos\varphi+\left(b_0+\delta b_1+\delta^2 b_2\right)\left(3\cos^2\varphi-1\right)\sin\varphi+O\left(\delta^3\right),\\
           \frac{\partial \left\langle\psi^{(2)}\right\rangle}{\partial \eta}\bigg|_{\eta = 0}&=\left(\delta c_1+\delta^2 c_2\right)\sin^2\varphi+\left(\delta d_1+\delta^2 d_2\right)\sin^2\varphi\cos\varphi+O\left(\delta^3\right),
         \endaligned
        \end{align}
where $a_0, a_1, a_2, b_0,b_1, b_2, c_1, c_2,d_1,d_2$ are complex constants that are functions of $V_0$ and $V_1$. Combining \eqref{eq16} and \eqref{eq17}, the inner flow can be solved order by order.
We will match the inner solution to the outer solution. As $\eta$ increases to infinity, the solution behaves as follows:
        \begin{align}\label{eq18}
         \aligned
\left\langle\psi^{(2)}\right\rangle&=\left[G_0+G_1\eta+G_2\eta^2+G_3\eta^3\right]\sin^2\varphi+\left(H_0+H_1\eta+H_2\eta^2+H_3\eta^3\right)\sin^2\varphi\cos\varphi,
         \endaligned
        \end{align}
The constants,  $G_i,H_i,\: i = 0,1,2,3$, can be expanded as orders of $\delta$:
\begin{align}\label{eq19}
G_i = G_{i0}+\delta G_{i1}+\delta^2 G_{i2}+O\left(\delta^3\right), \:
H_i = H_{i0}+\delta H_{i1}+\delta^2 H_{i2}+O\left(\delta^3\right).
\end{align}

In the outer region where $r \gg 1$, by substituting \eqref{eq13} into the right-hand side of \eqref{eq15}, we find it decays rapidly, on the order of $r^{-2}e^{-r}$, due to the appearance of the modified Bessel functions and their derivatives. Consequently, we can neglect these terms and solve the resulting homogeneous equation:
        \begin{align}\label{eq20}
         \aligned
           &D^4\left\langle \psi^{(2)}\right\rangle=0.
         \endaligned
        \end{align}
Considering the form of the inner flow in \eqref{eq18} and applying separation of variables to \eqref{eq20}, the outer flow takes the following form:
        \begin{align}\label{eq21}
         \aligned
           &\left\langle \psi^{(2)}\right\rangle=\left(T_1r^{-1}+T_2r\right)\sin^2\varphi+\left(T_3r^{-2}+T_4\right)\sin^2\varphi\cos\varphi,
         \endaligned
        \end{align}
where $T_i$ ($i = 1, 2, 3, 4$) are constants to be determined. Note that we have already applied the boundary condition that the fluid velocity decays to zero at infinity. To match \eqref{eq21} to \eqref{eq18}, we substitute $r = 1 + \delta \eta$ and expand \eqref{eq21} in orders of $\delta$. The constants $T_i$ are also expanded as follows:
\begin{align} \label{eq22}
T_i = T_{i0}+\delta T_{i1}+\delta^2 T_{i2}+O(\delta^3).
\end{align}
Comparing the coefficients in \eqref{eq19} and \eqref{eq22} to the order of $\delta$, we have
\begin{align}
    G_{00} &= T_{10}+T_{20}, G_{01} = T_{11}+T_{21},G_{11} = -T_{10}+T_{20},G_{12} = -T_{11}+T_{21},\\
    H_{00} &= T_{30}+T_{40}, H_{01} = T_{31}+T_{41},H_{11} = -2T_{30},H_{12} = -2T_{31}.
\end{align}
In the derivations above, the complex constants,$C_1, C_2, D_1, D_2, E_2$,  $a_0, a_1, a_2, b_0, b_1, b_2, c_1, c_2,$  $d_1, d_2$, $G_{00}, G_{01}, G_{11}, G_{12}$, $H_{00}, H_{01}, H_{11}, H_{12}$, are given in the Appendix. In the previous derivation, it is readily apparent that the flow field generated by oscillating bubble exhibits the characteristic scale of $\omega a$ before conducting nondimensionalization.
Consequently, the dimensional form of the streaming flow can be written as
        \begin{align}\label{eq27}
        \aligned
        &u_r=\epsilon^2\delta\omega a \bar{u}_r,\quad u_{\varphi}=\epsilon^2\delta\omega a \bar{u}_{\varphi},\\
        &\bar{u}_r=2\left(T_{1}r^{-3}+T_{2} r^{-1}\right)\delta^{-1}\cos\varphi+ \left(T_{3}r^{-4}+T_{4} r^{-2}\right)\delta^{-1}(3\cos^2\varphi-1),\\
        &\bar{u}_{\varphi}=\left(T_{1} r^{-3}-T_{2} r^{-1}\right)\delta^{-1}\sin\varphi+ 2T_{3}  r^{-4}\delta^{-1}\sin\varphi\cos\varphi.
        \endaligned
        \end{align}
As previously mentioned, we assume $\epsilon \ll \delta \ll 1$. Thus, we set $V_0=0.4$ and $V_1=1$ to calculate a representative acoustic streaming flow field. The resulting flow field is illustrated in the figure \ref{fig2}($a$), where the color-coded background represents the magnitude of stream function $\psi$ and the streamlines indicate the flow direction. The steady acoustic streaming flow is characterized by an axisymmetric double vortex structure. Under the same parameter settings, we calculate the time evolution results of the boundary deformation $R$ and $\Phi$ over one period, as illustrated in figure \ref{fig2}($b$)-($c$). The results indicate that the bubble boundary undergoes in-phase oscillations, exhibiting periodic deformation behaviors. In this case, the fluid surrounding the bubble is pushed
away along the $z$ axis and flows in from the $x$ direction, leading to the occurrence of vortex structure. Additionally, from equation \eqref{eq01}, it can be deduced that the bubble undergoes in-phase oscillations only when $V_0$ and $V_1$ are taken as real numbers. In this case, the coefficients  $T_{i0}$($i=1,2,3,4$) of stream function $\psi$ are purely imaginary, and the streaming flow is therefore generated at the order of $\epsilon^2\delta$. However, When either $V_0$ or $V_1$ is a complex number, the bubble undergoes out-of-phase oscillations, resulting in the leading order of the streaming flow being $O(\epsilon^2)$. Additionally, we select the same $V_0$ and $V_1$ to calculate the streaming flow at orders $\epsilon^2$ and $\epsilon^2\delta$, which are provided in the appendix.
We find that the streaming flow of order $\epsilon^2$ is characterized as a Stokeslet \citep{longuet1998}, while vortex structures are generated at order $\epsilon^2\delta$.  In our following investigations, we choose $V_0=0.4$ and $V_1=1$ to examine the impact of vortex generated by the in-phase oscillation of the bubble on particles in the acoustic streaming flow.

\subsection{Motion of an active particle}
In this section, we focus on the effect of acoustic streaming flow on the motion of particle within a two-dimensional framework. We assume that microorganisms respond instantaneously to the time-averaged flow field because of the significant time-scale separation between the time-averaging process and the swimming dynamics of the microorganisms. Specifically, the time-scale associated with the bubble oscillations that determine the time-averaged flow field is $O\left(10^{-6}\right)\sim O\left(10^{-4}\right)\si{s}$ \citep{yazdi2012}, while the time-scale of microorganism motion is $O\left(10^{-1}\right)\sim O\left(1\right)\si{s}$ \citep{wadhwa2022}. We introduce a dimensionless quantity $\alpha=\epsilon^2\delta\omega a/v_0$, which represents the ratio of characteristic flow strength $\epsilon^2\delta\omega a$, to the intrinsic swimming speed $v_0$. We rescale \eqref{eq08} and express it in component form as follows:
        \begin{align}\label{eq32}
        \aligned
             \frac{dr}{dt}&=\cos(\theta-\varphi)+\alpha \bar{u}_r,\\
             \frac{d\varphi}{dt}&=\frac{1}{r}\sin(\theta-\varphi)+\frac{\alpha \bar{u}_{\varphi}}{r},\\
             \frac{d\theta}{dt}&=\frac{\alpha}{2r}\left(\frac{\partial (r\bar{u}_{\varphi})}{\partial r}-\frac{\partial \bar{u}_r}{\partial \varphi}\right)+\frac{\alpha\gamma}{2}\left[\sin\left(2(\theta-\varphi)\right)\left(-\frac{\partial \bar{u}_r}{\partial r}+\frac{1}{r}\frac{\partial \bar{u}_{\varphi}}{\partial \varphi}+\frac{\bar{u}_{r}}{r}\right)\right.\\&\left.+\cos(2(\theta-\varphi))\left(r\frac{\partial}{\partial r}\left(\frac{\bar{u}_{\varphi}}{r}\right)+\frac{1}{r}\frac{\partial \bar{u}_r}{\partial \varphi}\right)\right].
        \endaligned
        \end{align}
Here, ($r$, $\varphi$) denotes the position of active particle in the flow field. Hence, we obtained a deterministic model that controls the behavior of particle motion in the flow field. The terms on the right-hand side of the equation \eqref{eq32} represent the deterministic rate of change with respect to $r$, $\varphi$ and $\theta$. Based on our estimates,  choosing $\epsilon = O\left(10^{-2}\right)\sim O\left(10^{-1}\right)$, $\delta = O\left(10^{-1}\right)\sim O(1)$, and $\omega a = O\left(10^5\right) \sim O\left(10^7\right)\si{\micro\meter/s}$ together with $v_0=O(10)\sim O\left(100\right)\si{\micro\meter/s}$ leads to a wide range of $\alpha$ values $\left(O\left(10^{-1}\right)\sim O\left(10^{3}\right)\right)$. The presence of curved streamlines in the acoustic streaming flow generates additional torques on particles, causing them to intricately align with local streamlines and modify their swimming behaviors. Additionally, the inherent activity of active particles can significantly influence the selection of their swimming direction.

\section{Impact of rotational diffusion}
\subsection{Discrete description}
Using a deterministic model \eqref{eq32}, we have developed an understanding of the swimming dynamics of a single particle in the acoustic streaming flow. However, in real biological environments, many factors, such as random walk behavior and thermal fluctuations, can influence the movements and distributions of microorganisms. Randomness plays a crucial role in the spatial heterogeneity of microorganisms. To gain insights into the effects of random fluctuations, it is necessary to incorporate noise into our deterministic model. Generally, noise primarily affects the motion behaviors of particles by inducing random translational and rotational diffusion. For a swimmer with typical size, rotational diffusion has more significant effects on the its movement compared to translational diffusion \citep{stark2016}. Here, we consider the rotational diffusion coefficient $D_r$ in our model to represent the random noise in the particle's orientation. To characterize the impact of diffusion on the swimming dynamics of particles, we define the Péclet number as:
         \begin{align}\label{eq34}
         \aligned
             &\Pen=\frac{v_0}{D_r a}.
         \endaligned
        \end{align}
We define $\mu_r$, $\mu_{\varphi}$, and $\mu_{\theta}$ as drift terms, representing the deterministic rates of change of $r$, $\varphi$, and $\theta$ in equation \eqref{eq32}. The Langevin equation for a particle, incorporating the effect of random rotational diffusion on its orientation, can be written as follows:
        \begin{align}\label{eq35}
        \aligned
             \frac{d\theta}{dt}&=\mu_{\theta}+\sqrt{2\Pen^{-1}}\xi_r(t),
        \endaligned
        \end{align}
where $\xi_r(t)$ is a random variable that satisfies $\langle \xi_r(t) \xi_r(t') \rangle = \delta_r(t - t')$, and $\delta_r$ follows the von Mises distribution. As a reference, for most microorganisms, the magnitude of the diffusivity is about $O\left(10^{-2}\right)\sim O\left(10^{-1}\right)\si{s^{-1}}$ \citep{Butenko2012, Koens2014, Tavaddod2011, Drescher2011}. As previously mentioned, the velocities of most microorganisms are  $O(10)\sim O\left(100\right) \si{\micro\meter/s}$. By setting $a \sim 10 \,\si{\micro\meter}$, the inverse Péclet number is $O\left(10^{-3}\right)$ to $O\left(10^{-1}\right)$.

\subsection{Continuous description}
To describe the statistical characteristics of the particles in the acoustic streaming flow, we introduce the probability distribution function $P(r,\varphi,\theta,t)$, which represents the probability that the particle is located at position ($r,\varphi$) and moving towards $\theta$ at time $t$. The probability distribution function is non-negative everywhere and its total integral is normalised to $1$. $P$ can be obtained by the Langevin equation\eqref{eq35}. We define an independent Wiener process $W$, satisfying $dW=\xi_r(t)dt$, where the increment $dW$ has zero mean and variance $dt$. Therefore, equation \eqref{eq35} can be rewritten as a stochastic differential equation (SDE),
        \begin{align}\label{eq36}
        \aligned
             d\theta&=\mu_{\theta}dt+\sqrt{2\Pen^{-1}}dW,
        \endaligned
        \end{align}
where increment $dW$ satisfies
        \begin{align}\label{eq37}
        \aligned
        &\langle dW\rangle=0,\quad (dW)^2=dt.
        \endaligned
        \end{align}
Then, we can obtain the \textit{multidimensional It$\hat{o}$ formula} from the stochastic differential equation \eqref{eq36}. This formula can be used to transform the stochastic process into an evolution process of probability density.
We therefore derive the corresponding Smoluchowski equation as follows:
        \begin{align}\label{eq45}
         \aligned
            \frac{\partial P}{\partial t}&=-\frac{\partial }{\partial r}\left(\mu_rP\right)-\frac{\partial}{\partial \varphi}\left(\mu_{\varphi}P\right)-\frac{\partial}{\partial \theta}\left(\mu_{\theta}P\right)+\Pen^{-1}\frac{\partial^2P}{\partial \theta^2},
         \endaligned
        \end{align}
where
        \begin{align}\label{eq46}
        \aligned
             \mu_r&=\cos(\theta-\varphi)+\alpha \bar{u}_r,\\
             \mu_{\varphi}&=\frac{1}{r}\sin(\theta-\varphi)+\frac{\alpha \bar{u}_{\varphi}}{r},\\
             \mu_{\theta}&=\frac{\alpha}{2r}\left(\frac{\partial (r\bar{u}_{\varphi})}{\partial r}-\frac{\partial \bar{u}_r}{\partial \varphi}\right)+\frac{\alpha\gamma}{2}\left[\sin\left(2(\theta-\varphi)\right)\left(-\frac{\partial \bar{u}_r}{\partial r}+\frac{1}{r}\frac{\partial \bar{u}_{\varphi}}{\partial \varphi}+\frac{\bar{u}_{r}}{r}\right)\right.\\&\left.+\cos(2(\theta-\varphi))\left(r\frac{\partial}{\partial r}\left(\frac{\bar{u}_{\varphi}}{r}\right)+\frac{1}{r}\frac{\partial \bar{u}_r}{\partial \varphi}\right)\right].
        \endaligned
        \end{align}

\section{Results}
\subsection{Motion of a single particle}

\begin{figure}
           \centerline{
            \includegraphics[width=13.0cm]{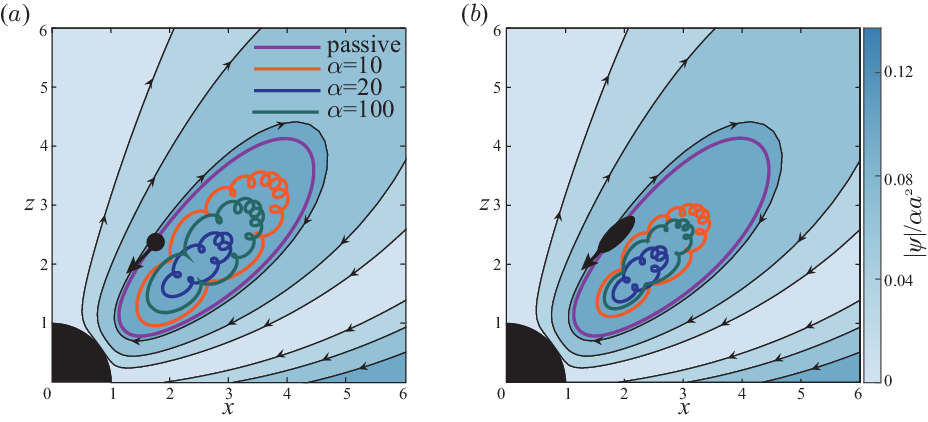}}
            \captionsetup{justification=justified, singlelinecheck=false}
            \caption{The stable bound orbits of
            passive and active particles : ($a$) spherical ($\gamma=0$), ($b$) elongated ($\gamma=1$). We consider different flow strengths, $\alpha=10$, $20$, $100$ in different colors. The initial position of each particle is $(r_0, \varphi_0)=(3, \pi/5)$ (black circular and elliptical point), and the initial orientation angle is $\theta_0=5\pi/4$ (black arrows). The magnitude of the scaled stream function is $|\psi|/\alpha a^2$.}
             \label{fig3}
     \end{figure}

We investigate the swimming dynamics of both passive and active particles under identical initial conditions in the acoustic streaming flow. 
For the numerical results, we solve equation \eqref{eq32} using a fourth-order Runge-Kutta method, as employed in the study by \citet{jos2020}, ensuring a sufficiently small time interval to maintain stability and accuracy. The particles are initially positioned at $r_0=3$, $\varphi_0=\pi/5$, with an orientation angle of $\theta_0=5\pi/4$. Upon exposure to the streaming flow, these particles tend to align with the streamlines. Over time, their positions and orientations exhibit periodic variations, indicating that they are trapped in stable bounded orbits. Figure \ref{fig3}$(a)$ demonstrates the stable bounded orbits for a spherical particle ($\gamma=0$) at different flow strengths $\alpha$. We observe that the passive particle moves along the streamline passing through its initial position, whereas the stable bounded orbit of the active particle is closer to the vortex center and has a smaller range compared to that of passive particle. In this context, the range of a stable bounded orbit refers to the extent of variation in both the radial and polar directions of the orbit, describing its spatial coverage. This difference of orbit indicates that the inherent activity of active particles enables them to cross streamlines to enter the region near the vortex center, leading to a reduction in the range of their stable bounded orbits.

Additionally, we observe that the stable bounded orbit of an active particle depends on the flow strength $\alpha$. When $\alpha=10$, the active particle crosses the streamlines and is trapped in the region near the vortex center. As $\alpha$ increases to $20$, the stable bounded orbit becomes closer to the vortex center compared to the case of $\alpha=10$. However, for $\alpha=100$, the range of stable bounded orbits for the active particles increases compared to the case of $\alpha=20$. In this context, the acoustic streaming flow dominates the behavior of the active particle. As the particle moves into regions far from the oscillating bubble, the cross-flow velocity of the streaming flow aligns it with the streamlines, reducing its ability to enter the interior of the vortex, and ultimately trapping it in the stable bounded orbit farther from the vortex center.

In figure \ref{fig3}$(b)$, we demonstrate the stable bounded orbits of elongated particles ($\gamma=1$) under the same initial conditions. We find that in both spherical and elongated cases, passive particles move along the streamlines. Elongated active particles are also trapped in the stable bounded orbits near the vortex center similar to the spherical cases. However, at the given flow strength $\alpha$ and initial conditions, the range of the stable bounded orbit of an elongated active particle is systematically smaller than that of a spherical particle. This could be due to the fact that elongated particles are more easily able to cross streamlines. According to Jeffery’s orbit \citep{jeffery1922}, elongated particles exhibit more significant and complex rotational behaviors compared to spherical particles. Specifically, when elongated particles are in motion, their rotations tend to follow Jeffery’s orbit, which results in greater alignment with the flow direction and facilitates their ability to move across streamlines. Therefore, it is reasonable to expect that elongated particles, when they have velocity, will have a greater tendency to cross streamlines, leading to a smaller overall motion range. On the other hand, we hypothesize that elongated particles may migrate toward regions with higher shear rates, particularly near the vortex center. This is similar to the findings of \citet{Nitsche1997}, who reported that elongated particles in shear flows migrate toward regions with higher shear rates, such as near the wall. However, due to the complexity of the flow field in our study, the exact cause of this migration cannot be fully confirmed.
           \begin{figure}
           \centerline{
            \includegraphics[width=13.0cm]{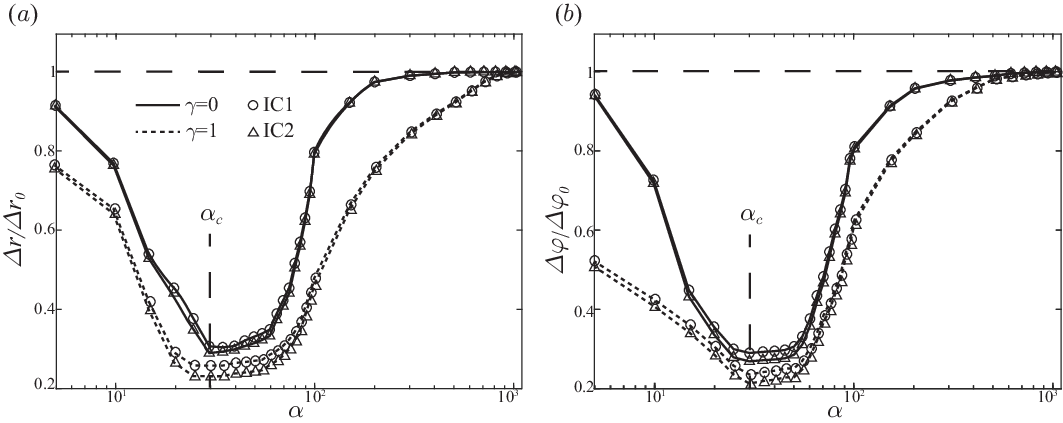}}
            \captionsetup{justification=justified, singlelinecheck=false}
            \caption{Dependence of the stable motion range of active particle on the flow strength $\alpha$. Motion ranges are measured by ($a$) $\Delta r$ along the radial direction and ($b$) $\Delta \varphi$ along the polar direction, respectively. These ranges are normalized by the $\Delta r_0$ and $\Delta \varphi_0$, which represents the motion range of the passive particle along the radial and polar direction, respectively.
            We consider the spherical ($\gamma=0$) and elongated ($\gamma=1$) active particles under two different initial condition: IC1($r_0=3$, $\varphi_0=\pi/5$, $\theta_0=5\pi/4$) and IC2($r_0=4$, $\varphi_0=\pi/4$, $\theta_0=\pi/4$).}
             \label{fig4}
           \end{figure}

The stable bounded orbit of an active particle depends on the flow strength $\alpha$ and the shape parameter $\gamma$, indicating that the interaction between the background flow field and the particle can significantly influence the particle's range of motion. We further investigate the dependence of the stable motion ranges of both spherical and elongated active particles on flow strength $\alpha$. We use $\Delta r$ and $\Delta \varphi$ to represent the motion range of particles along the radial and polar directions, respectively, when the dynamics reach a steady state. Specifically, once the motion reaches a steady periodic state, we extract the maximum and minimum values of $r$ and $\varphi$ during the periodic motion and calculate their differences to obtain $\Delta r$ and $\Delta \varphi$. These ranges are then normalized by the corresponding values for passive particles, denoted by $\Delta r_0$ and $\Delta \varphi_0$. To ensure consistency across all cases, we use identical initial conditions, which eliminates the influence of initial conditions on the results. In figure \ref{fig4}, we present the stable motion ranges $\Delta r$ and $\Delta \varphi$ of spherical ($\gamma=0$) and elongated ($\gamma=1$) active particles under different flow strengths $\alpha$. We consider two initial conditions: the first case $(r_0,\varphi_0,\theta_0)=(3,\pi/5,5\pi/4)$ (IC1) and the second case $(r_0,\varphi_0,\theta_0)=(4,\pi/4,\pi/4)$ (IC2). We note that although we present results for two different initial conditions (IC1 and IC2), the ratio of motion ranges between active and passive particles remains qualitatively similar in both cases, indicating that the observed trends are not sensitive to the initial conditions. We observe that as the flow strength $\alpha$ increases, the motion ranges $\Delta r$ and $\Delta \varphi$ of the particles initially decrease for all cases. When the flow strength $\alpha$ reaches a critical value of $30$, the motion ranges $\Delta r$ and $\Delta \varphi$ of active particles are minimized. As the flow strength $\alpha$ continues to increase, the stable motion range of the active particle increases until it converges to $1$.

This non-monotonic behavior arises from the interaction between the intrinsic activity of the active particles and the external acoustic streaming flow. At low flow strengths, the particle’s intrinsic velocity dominates its motion, and its orientation is random, leading to a relatively broad motion range. At very high flow strengths, the particle's intrinsic velocity becomes much smaller than the flow velocity, causing its behavior to closely resemble that of a passive particle. In this regime, the particle's motion is restricted to follow the streamlines, with its behavior determined by its initial conditions. However, when the velocity scales of the particle and the flow become comparable, the particle exhibits more intricate dynamics, such as crossing streamlines and remaining confined nearer to the vortex center. Consequently, the motion range is relative large at both low and high flow strengths, resulting in the observed non-monotonic dependence of motion range on flow strength.

Additionally, we find that the stable motion range  $\Delta r$ and $\Delta \varphi$ for active particles is nearly identical under different initial conditions. This indicates that the initial position and orientation of the particle have little effect on its motion range. This indicates that the initial position and orientation of the particle have little effect on its motion range. Moreover, under the given flow strength $\alpha$ and initial conditions, the stable motion range of an elongated particle is smaller than that of spherical particle. This further indicates that the elongated geometry of the particle enhances its ability to cross streamlines and enter the region near the vortex center.Therefore, we demonstrate that the interaction between the flow field and active particles has a crucial impact on the stable bounded orbits and motion ranges of active particles.

\subsection{Dependence of trapping on the flow strength}

We investigate the trapping of passive and active particles by the acoustic streaming flow in the presence of rotational diffusion. In the deterministic model, the stable bounded orbits of spherical and elongated particles are qualitatively similar. Therefore, we primarily focus on the case of an elongated particle in the following analysis. We initialize 1500 particles uniformly distributed within the region $1\leqslant r\leqslant 10$ and $0\leqslant \varphi\leqslant \pi/2$, with their initial orientation angle set randomly within the range ($0,2\pi$). Under the influence of the acoustic streaming flow, the particles exhibit two possible behaviors: escaping from the flow field or being trapped within the streaming flow field. Particles are considered to be trapped when their motion stabilizes into a bounded orbit within the computational domain, indicating confinement by the streaming flow. However, particles are considered to escape when they cross the boundaries $r=1$, $r=10$, $\varphi=0$ and $\varphi=\pi/2$. To account for this, we apply the absorbing boundary conditions to remove particles that reach the boundaries $r=1$, $r=10$, $\varphi=0$ and $\varphi=\pi/2$.  We numerically solve the equation (\ref{eq35}) by using the Euler-Maruyama method to calculate the trajectories until the particles either escape or their motion ranges stabilize. Thus, we can obtain the stable distribution positions of the trapped particles under different flow strengths. 
           \begin{figure}
           \centerline{
            \includegraphics[width=13.0cm]{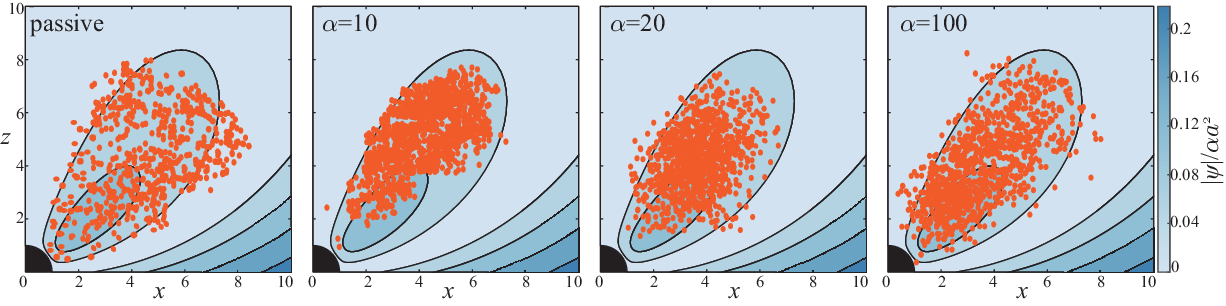}}
            \captionsetup{justification=justified, singlelinecheck=false}
            \caption{The stable distribution of trapped passive and active particles represented by orange points. We consider different flow strength $\alpha=10$, $\alpha=20$, $\alpha=100$. The corresponding parameters are $\Pen^{-1}=0.01$ and $\gamma=1$. The magnitude of the scaled stream function is $|\psi|/\alpha a^2$.}
             \label{fig5}
           \end{figure}

Figure \ref{fig5} illustrates the stable distribution of the trapped passive and active particles in the acoustic streaming flow. The orange points shown in figure \ref{fig5} correspond to the instantaneous positions of individual particles after the system has reached a statistically steady state, providing a discrete visualization that complements the continuum probability density functions.
We observe that passive particles are distributed dispersedly within the region of closed streamlines. In contrast, active particles exhibit a tendency to aggregate in specific regions of the flow field, with their stable distribution area being significantly smaller than that of passive particles. When $\alpha=10$, the distribution region of the particles is relatively confined, with active particles aggregating in regions far from the oscillating bubble. When $\alpha=20$, active particles are primarily distributed in the region near the vortex center.  As the flow strength $\alpha$ increases further, reaching $\alpha=100$, the distribution region expands and becomes more dispersed compared to the case at $\alpha=20$. These observations indicates that the stable distribution region of active particles depends on the flow strength.

           \begin{figure}
           \centerline{
            \includegraphics[width=8.0cm]{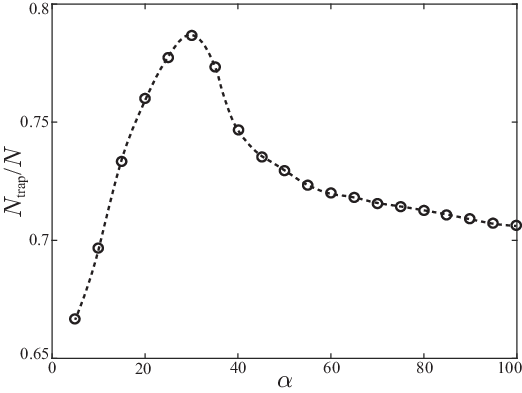}}
            \captionsetup{justification=justified, singlelinecheck=false}
            \caption{Dependence of trapping rate $N_{\rm trap}/N$ of active elongated particles ($\gamma=1$) on the flow strength. $N_{\rm trap}$ is the number of trapped particles while $N$ is the total number of particles initially located in the streaming flow.}
             \label{fig6}
           \end{figure}

Then, we define the trapping rate as the proportion of active particles that remain confined within the computational domain after their motion stabilizes in a steady state. To calculate this, the positions of all particles are tracked throughout the simulation, and those that remain within the domain are identified. The trapping rate is then determined as the ratio of the number of confined particles, $N_{\rm trap}$, to the total number of particles initially introduced into the computational domain, $N$. In figure \ref{fig6}, we present the trapping rate of active particles under different flow strengths. We observe that the trapping rate of active particles initially increases with the flow strength $\alpha$, as the particle's intrinsic activity allows for stable confinement within the vortex region. When $\alpha$ reaches a critical value of $30$, the number of trapped active particles reaches its maximum, with the majority of active particles being confined within the acoustic streaming flow field. When the flow strength $\alpha$ exceeds the critical value and continues to increase, the number of active particles trapped in the acoustic streaming flow gradually decreases. As previously mentioned in figure \ref{fig4}, at low flow strength $\alpha$, the intrinsic activity of particle dominates its motion, leading to a larger motion range, which increases the likelihood of particles escaping under the influence of rotational diffusion, resulting in a lower trapping rate. As $\alpha$ increases to intermediate values, the flow strength becomes comparable to particle activity, reducing their motion range and confining them closer to the vortex center. This confinement minimizes escape probabilities and maximizes the trapping rate. However, at high flow strengths $\alpha$, the flow field dominates the particle's behavior, and the motion range increases again, particularly near the vortex boundary. The increased motion range, combined with rotational diffusion, results in a higher probability of escape, causing the trapping rate to decline.

\subsection{Distribution of the trapped particles}

We further investigate the distribution of trapped passive and active particles under different flow strengths. On the one hand, following a same approach as previously described, we initialize 1500 particles uniformly distributed within the region $1\leqslant r\leqslant 10$ and $0\leqslant \varphi\leqslant \pi/2$, with their initial orientation angle $\theta$ set randomly within the range ($0,2\pi$). The boundaries of this calculation region, defined by $r=1$, $r=10$, $\varphi=0$ and $\varphi=\pi/2$, are set as absorbing to remove particles that reach the boundaries. Then, we calculate the distribution positions of all trapped particles by solving the Langevin equation until the particle distributions converge over time, indicating that a steady-state has been reached. Subsequently, the calculation domain is divided into a uniform grid of $200$ cells in the $r$ direction and $100$ cells in the $\varphi$ direction. We determine the number of particles in each grid cell and normalize these counts by the total number of particles. By averaging the normalized counts over the angular direction for each radial segment and over the radial direction for each angular segment, we can obtain stable radial distribution function (RDF) and angular distribution function (ADF) of particles.

On the other hand, we numerically solve the Smoluchowski equation (\ref{eq45}) by employing the Crank-Nicolson method. This computation is performed within the region defined by $1\leqslant r\leqslant 10$, $0\leqslant \varphi\leqslant \pi/2$ and  $0\leqslant \theta\leqslant 2\pi$. The initial probability distribution function $P$ within the calculation region has a uniform distribution that satisfies the normalization condition. To maintain consistency with the boundary conditions employed in the Langevin equation, we impose absorbing boundary conditions with $P=0$ at the boundaries $r=1$, $r=10$, $\varphi=0$ and $\varphi=\pi/2$.  Additionally, a periodic condition is imposed to $\theta$. The numerical solution is carried out on a mesh grid of dimensions
 $200\times 100\times400$, providing sufficient resolution to capture the distribution's dynamics. The solution is iteratively computed until a steady-state probability distribution function $P_s$ is obtained, with convergence monitored throughout the process to ensure accuracy. Therefore, the stable radial distribution function (RDF) and angular distribution function (ADF) of particles can be calculated by
        \begin{align}\label{eq50}
         \aligned
            &{\rm RDF}(r)=\iint P_s\, d\varphi d\theta,\quad
            {\rm ADF}(\varphi)=\iint P_s r\, drd\theta.
         \endaligned
        \end{align}

          \begin{figure}
           \centerline{
            \includegraphics[width=13.0cm]{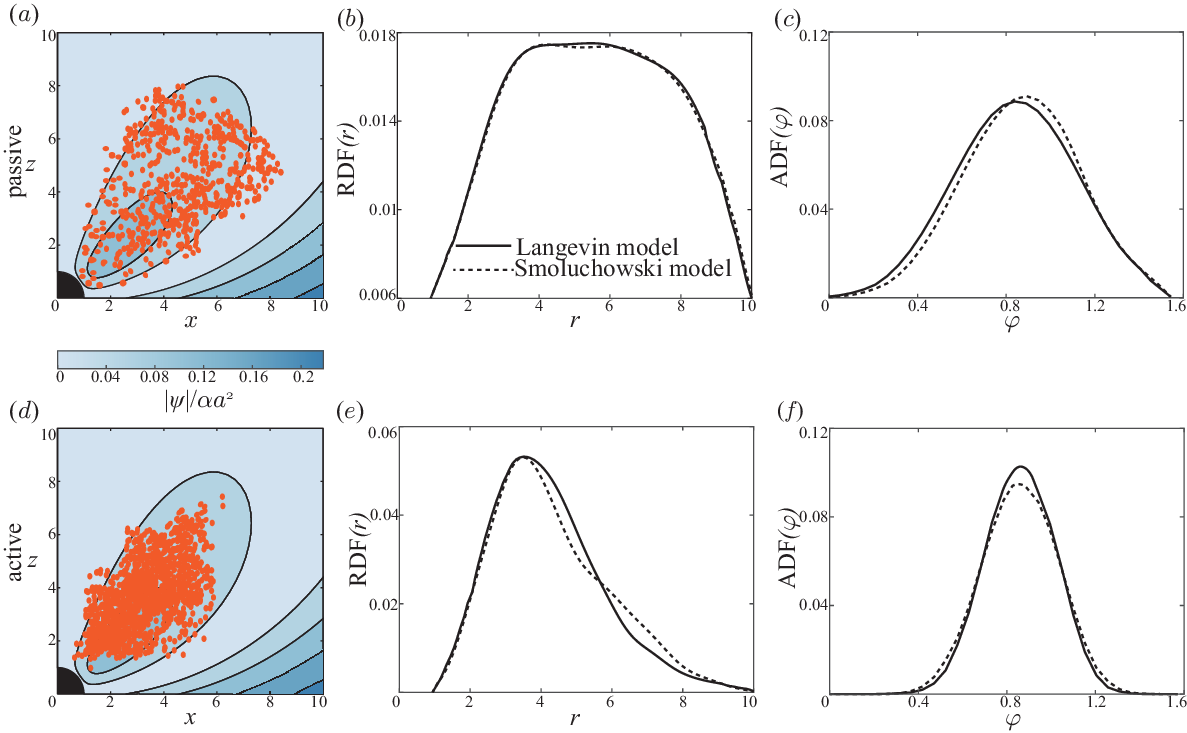}}
            \captionsetup{justification=justified, singlelinecheck=false}
            \caption{The stable distribution of  ($a$) passive and ($d$) active particles obtained from the Langevin model. The distribution of particles are determined by using radial (RDF) and angular distribution functions (ADF): ($b-c$) passive particles and ($e-f$) active particles, which are compared with results obtained from Smoluchowski equation. The corresponding parameters are $\alpha=60$, $\Pen^{-1}=0.01$ and $\gamma=1$.}
             \label{fig7}
           \end{figure}

 Figure \ref{fig7} demonstrates the stable distribution of the trapped passive and active particles at $\alpha=60$, $\Pen^{-1}=0.01$ and $\gamma=1$. Additionally, in figure \ref{fig7} ($b$)-($c$) and ($e$)-($f$), we compare the stable radial distribution function (RDF) and angular distribution function (ADF) of these particles obtained from the Langevin model \eqref{eq35} and the Smoluchowski equation \eqref{eq45}. The consistency between the results from both models confirms the accuracy and effectiveness of our approach. Figure \ref{fig7} ($a$)-($c$) shows that the distribution of passive particles is relatively uniform in the radial direction, with particles primarily distributed in regions of closed streamlines in the angular direction. Passive particles initially located on unclosed streamlines escape from the flow field due to the incoming flow, whereas only particles located on the closed streamlines can remain confined within the flow field and continue moving along streamlines because of their lack of activity. Thus, passive particles are uniformly distributed in the region composed of closed streamlines. However, in figure \ref{fig7} ($d$)-($f$), active particles exhibit a significant concentration in the central region of the vortex, leading to a non-uniform distribution in both the radial and angular direction. Unlike passive particles, active particles can cross the streamlines laterally and enter the region near the vortex center, regardless of whether their initial positions are on the enclosed streamlines. Therefore, active particles located within the acoustic streaming flow region can be trapped and collected in the smaller region near the vortex center compared with that of passive particles, resulting in a non-uniform distribution.

           \begin{figure}
           \centerline{
            \includegraphics[width=13.0cm]{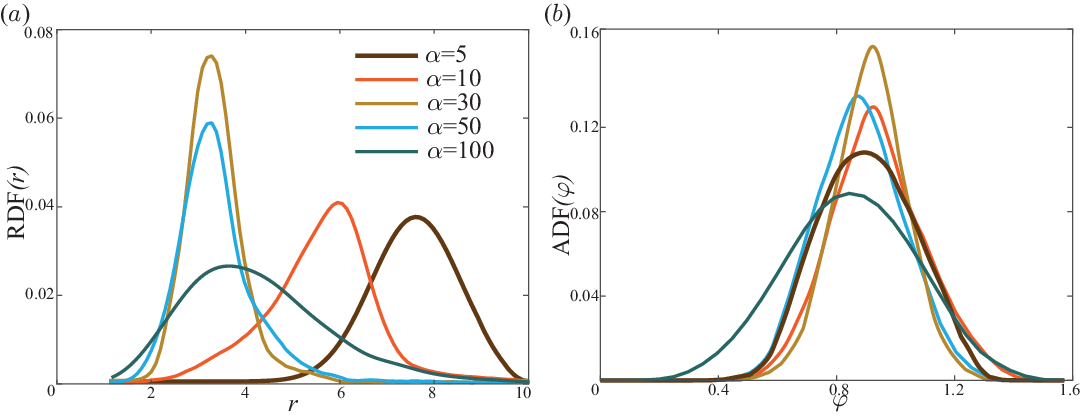}}
            \captionsetup{justification=justified, singlelinecheck=false}
            \caption{(color online) The stable ($a$) radial distribution function (RDF) and ($b$) angular distribution function (ADF) of active particles. We consider different flow strength $\alpha=5,10,30,50,100$ in different colors. Other corresponding parameters are $\gamma=1$, $\Pen^{-1}=0.01$.}
             \label{fig8}
           \end{figure}

To further elucidate the dependence of the stable distribution of active particles on the flow strength $\alpha$, we calculate the radial distribution function and angular distribution function of active particles under different flow strengths within the range $(1<\alpha<100)$.
In figure \ref{fig8},  we demonstrate the typical results of $\alpha=5$, $10$, $30$, $50$ and $100$. We observe that, at $\alpha=5$, active particles concentrate in the region further away from the vortex center in the radial direction. However, as the flow strength $\alpha$ continues to increase, the concentration position of active particles in the radial direction is closer to the vortex center, and the particles become more concentrated compared to the case of $\alpha=5$. When the flow strength reaches a critical value of $\alpha=30$, almost all trapped active particles are collected in the vortex center of the acoustic streaming flow field, where the degree of accumulation reaches its maximum in both radial and angular directions. Thus, for $\alpha\leq 30$, increasing the flow strength $\alpha$ enhances the accumulation degree of active particles. However, when $\alpha$ continues to increase, as in the cases of $\alpha=50, 100$, the accumulation degree of active particles decreases in both radial and angular directions. The accumulation behavior of active particles strongly depends on individual particle dynamics. At very low or high flow strengths, the particle’s motion range is large, leading to weak accumulation. When the flow is comparable to the particle's velocity, the motion range of particle becomes smaller, causing a more pronounced accumulation. Consequently, accumulation behavior of active particles shows a non-monotonic dependence on flow strength. Our results show that the interaction between the acoustic streaming flow and particle activity dominates the distribution position and accumulation degree of active particles.

\subsection{The effect of noise on the distribution of particles}

\begin{figure}
           \centerline{
            \includegraphics[width=13.0cm]{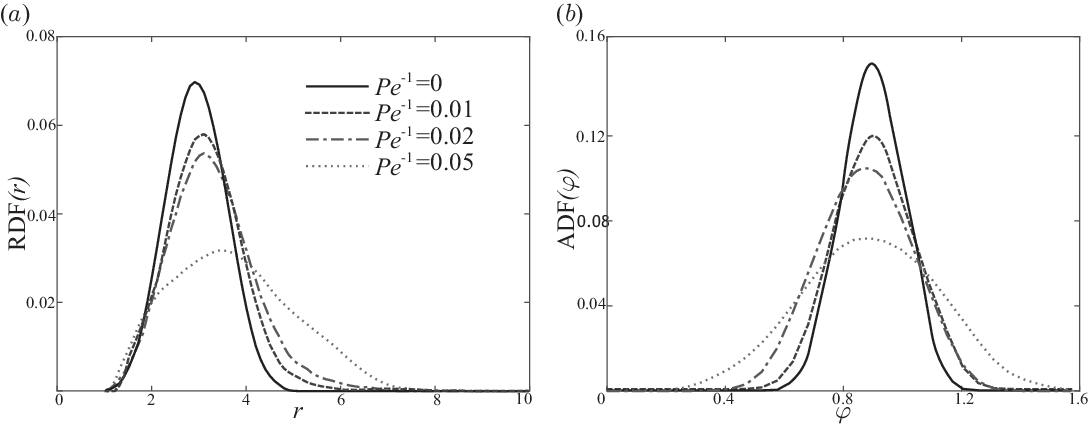}}
            \captionsetup{justification=justified, singlelinecheck=false}
            \caption{ The stable ($a$) radial distribution function and ($b$) angular distribution function of active particles at $\gamma=1$ and $\alpha=25$. We consider different Péclet number, $\Pen^{-1}=0, 0.01, 0.02, 0.05$.}
             \label{fig9}
           \end{figure}

Generally, rotational diffusion significantly affects the distribution of active particles. Thus, we calculate the radial distribution function and angular distribution function of active particles under $\Pen^{-1}=0$, $0.01$, $0.02$ and $0.05$.
Figure \ref{fig9} demonstrates the obtained results. We can observe that as $\Pen^{-1}$ increases, the aggregation degree of active particles in both radial and angular directions decreases. In other words, an increase in rotational diffusion leads to a more dispersed distribution of active particles under a given flow strength $\alpha$. This is because the enhanced rotational diffusion increases the randomness of their orientation, reducing the influence of the flow field on their directional movement. As a result, active particles are more likely to diffuse throughout the streaming flow field rather than being concentrated in specific regions, leading to a more dispersed overall distribution.

\section{Concluding remarks}

We investigate the swimming dynamics of both passive and active particles within the acoustic streaming flow generated by an oscillating bubble. Through the establishment of a deterministic model, we demonstrate the existence of stable bounded orbits for both passive and active particles in the absence of noise. Our findings indicate that passive particle can only moves along the streamline passing through its initial position, while active particle has the ability to cross streamlines and become trapped in stable bounded orbits closer the vortex center compared to that of passive particle. Then, we define the flow strength $\alpha$, which is the radio of characteristic flow strength of acoustic streaming flow to the intrinsic swimming speed of particles. We find that the flow strength dominates the stable bounded orbit of the active particle. Therefore, we further investigate the effect of flow strength $\alpha$ on the stable motion range of active particles. We observe that as the flow strength increases, the stable motion range of active particle initially decreases. However, after exceeding a critical flow strength, the range of particle motion gradually increases with $\alpha$ until it becomes identical to that of passive particle. This is attributed to the interaction between particle activity and the flow field. Furthermore, we demonstrate that the features of both the stable bounded orbits and the trends in their stable motion range variation with flow strength remain consistent for the spherical and slender shape particles. 

We further introduce rotational noise to examine how the trapping and distribution of passive and active particles are influenced by flow strength, utilizing the Langevin and Smoluchowski equations. We find that the distribution range and trapping rate of the active particles initially increases and then decreases with increasing flow strength $\alpha$, which is consistent with the observations of single particle. However, active particles exhibit a significant concentration in the central region of the vortex, leading to a non-uniform distribution in both the radial and angular direction. This phenomenon provides insight into the observed accumulation of \textit{Escherichia coli} in the acoustic streaming flow field \citep{yazdi2012}. Additionally, as the flow strength increases, the concentration region of the active particle is closer to the vortex center, while the accumulation degree of active particles initially increase and then decreases. Additionally, we identify a flow strength that maximizes the accumulation degree of active particles within the acoustic streaming flow. The particle accumulation strongly depends on the motion range of individual particle, resulting in the non-monotonic dependence on the flow strength, which represents a key insight from our study. This result has practical implications for microfluidic systems, suggesting that flow strength can serve as a control parameter to modulate particle behavior for targeted aggregation or dispersion.

We remark several limitations of the current study and explore potential avenues for future research. First, in our study, we assume both the motion of particles and the streaming flow field to be two-dimensional. This simplification facilitates a more straightforward investigation of particle dynamics in the acoustic streaming flow. However, allowing particles to move in the third dimension can potentially reveal new behaviors not observed in the two-dimensional case. The emergence of complex vortex structures in three-dimensional acoustic flows may lead to altered particle aggregation characteristics and the appearance of scattering behavior \citep{ahmed2016}. Second, our study considers the flow field and particle motion to be unbounded, whereas real-world microfluidic environments are typically bounded. In typical microfluidic conditions, oscillating microbubble is ususlly trapped within microfluidic chips, introducing additional boundary conditions such as wall-induced hydrodynamic interactions. These effects may alter the flow field and particle dynamics \citep{Dey2022}, which are not accounted for in our current model. Incorporating reflective or other boundary conditions with bounded flow fields, will provide a more comprehensive understanding of particle dynamics in acoustic streaming flows, which is essential for translating theoretical findings into practical applications in microfluidic devices. Third, our analysis is based on an asymptotic solution of the acoustic streaming flow, which requires $\epsilon \ll \delta \ll 1$. While this approximation provides a tractable way to capture the essential vortex structure and define a meaningful flow strength scale, it may not perfectly represent real experimental conditions, especially as $\epsilon$ and $\delta$ approach $O\left(10^{-1}\right)$ and $O(1)$, respectively. In such regimes, higher-order effects, boundary conditions, or other complexities may become significant, and more detailed numerical simulations or direct experimental measurements would be required to accurately quantify the flow field and particle dynamics. Finally, We have focused on the motion of non-gyrotactic particles with neglecting inertial effects. Future research can explore the role of inertia and gyrotaxis, especially in microorganisms like algae, to better understand particle behavior in acoustic streaming flows and its implications for biological and fluid systems.

 Our study provides valuable insights into the hydrodynamic mechanisms underlying the swimming and aggregation behaviors of active particles within acoustic streaming flow fields. These findings extend beyond basic research and have broader implications across various fields. For instance, the focusing mechanisms of particles under varying flow strengths, as demonstrated in our study, offer indirect evidence supporting the existence of microbial heterogeneity within the human body's environment \citep{weiss2023, she2024}, highlighting the intricate interdependence between microorganisms and their non-biological surroundings, particularly in biofilm formation. Moreover, the aggregation behavior of active particles in acoustic streaming flows can be harnessed for practical applications such as developing disposable and portable microfluidic devices for rapid microbial detection \citep{Rossi2023}. By optimizing flow strength, these devices can collect microorganisms more quickly to detectable levels, significantly advancing diagnostic technology. In environmental monitoring, similar mechanisms can be applied to efficiently detect pollutant particles in water and air, improving the speed and accuracy of pollution detection. Furthermore, the control of cell aggregation in tissue engineering and the development of high-precision laboratory equipment in microfluidics and acoustic manipulation can benefit from our insights, paving the way for innovations in areas like single-cell analysis and particle sorting.

\section*{Acknowledgements}
We would like to thank Prof. Eva Kanso and Dr. Zhiwei Peng for their invaluable discussions and  the anonymous reviewers for their valuable comments in improving this manuscript.

\section*{Fundings}
We acknowledge support from the National Key Research and Development Program of China No. 2021YFA1000200, 2021YFA1000201 (to W. Tan) and the National Natural Science Foundation of China under grant No. 12372258 (to Y. Man).
\section*{Declaration}
The authors report no conflict of interest.

\section*{Appendix A. Coefficients and flow characteristics}

In the section of sketching the asymptotic solution, the relevant complex constants,$C_1, C_2$, $D_1, D_2, E_2$,  $a_0, a_1, a_2, b_0, b_1, b_2, c_1, c_2,d_1,d_2$, $G_{00}, G_{01}, G_{11}, G_{12}$, $H_{00}, H_{01}, H_{11}, H_{12}$, are given as follows:
        \begin{align}
         \aligned
           &C_1=\frac{3\sqrt{2}}{4}(1+i)\bar{V}_0V_1,\quad C_{21}=\left(\frac{39}{2}-24\sqrt{2}\right)\bar{V}_0V_1,\quad C_{22}=-6(1+i)\bar{V}_0V_1,\\
           &D_1=\frac{3\sqrt{2}}{4}(1+i)\bar{V}_1V_1,\quad D_{21}=\left(-30+36\sqrt{2}\right)\bar{V}_1V_1,\quad D_{22}=-9(1+i)\bar{V}_1V_1,\\
           & E_2=3i\bar{V}_1V_1,\quad a_0=V_0\bar{V}_1i+\frac{3i}{2}\bar{V}_0V_1,\quad a_1=-\frac{3\sqrt{2}}{2}(1+i)\bar{V}_0V_1,
         \endaligned
        \end{align}

        \begin{align}
         \aligned
         &a_2=39\bar{V}_0V_1, \quad b_0=\frac{27i}{40}\bar{V}_1V_1,\quad b_1=-\frac{3\sqrt{2}}{4}(1+i)\bar{V}_1V_1,\quad b_2=(30-3i)\bar{V}_1V_1,\\
          &c_1=\frac{11i}{4}\bar{V}_0V_1+\frac{i}{2}V_0\bar{V}_1,\quad c_2=\left[-\frac{1}{4}+\left(6-\frac{39\sqrt{2}}{4}\right)(1+i)\right]\bar{V}_0V_1,\\
           &d_1=\frac{11i}{4}\bar{V}_1V_1,\quad d_2=\left[\left(9-\frac{27\sqrt{2}}{2}\right)(1+i)-\frac{27}{20}\right]\bar{V}_1V_1,\\
           &G_{00}=\frac{3i}{4}\bar{V}_0V_1+\frac{i}{2}V_0\bar{V}_1,\quad G_{01}=0, \quad G_{11}=\frac{5i}{4}\bar{V}_0V_1+\frac{i}{2} V_0\bar{V}_1,\quad G_{12}=-\frac{1}{4}\bar{V}_0V_1,\\
           &H_{00}=\frac{27i}{40}\bar{V}_1V_1,\quad H_{01}=0,\quad H_{11}=\frac{5i}{4}\bar{V}_1V_1, \quad H_{12}=-\frac{27}{20}\bar{V}_1V_1,\\
           &T_{10}=-\frac{i}{4}\bar{V}_0V_1,\quad T_{11}=\frac{1}{8}\bar{V}_0V_1,\quad T_{20}=\bar{V}_0V_1i+\frac{i}{2}V_0\bar{V}_1,\quad T_{21}=-\frac{1}{8}\bar{V}_0V_1,\\
           &T_{30}=-\frac{5i}{8}\bar{V}_1V_1,\quad T_{31}=\frac{27}{40}\bar{V}_1V_1,\quad T_{40}=\frac{13i}{10}\bar{V}_1V_1,\quad T_{41}=-\frac{27}{40}\bar{V}_1V_1.
         \endaligned
        \end{align}

\begin{figure}
           \centerline{
            \includegraphics[width=8.0cm]{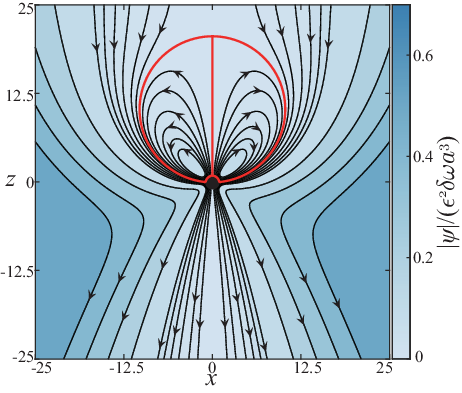}}
            \captionsetup{justification=justified, singlelinecheck=false}
            \caption{The steady acoustic streaming flow   generated by the oscillating bubble for $V_0=0.4$, $V_1=1$. The magnitude of the scaled stream function is $|\psi|/(\epsilon^2\delta\omega a^3$). The red line $\psi=0$ represents the separatrix, which separates the closed-streamline regions from the open-streamline regions.}
             \label{fig10}
           \end{figure}

In addition, we present the steady acoustic streaming flow with $V_0=0.4$ and $V_1=1$, which are used in our study for particle dynamics in figure \ref{fig10}. The separatrix ($\psi=0$), is shown as the boundary between the closed and open streamline regions.
\section*{Appendix B. Dependency of steady orbits on initial conditions }
\begin{figure}
           \centerline{
            \includegraphics[width=13.0cm]{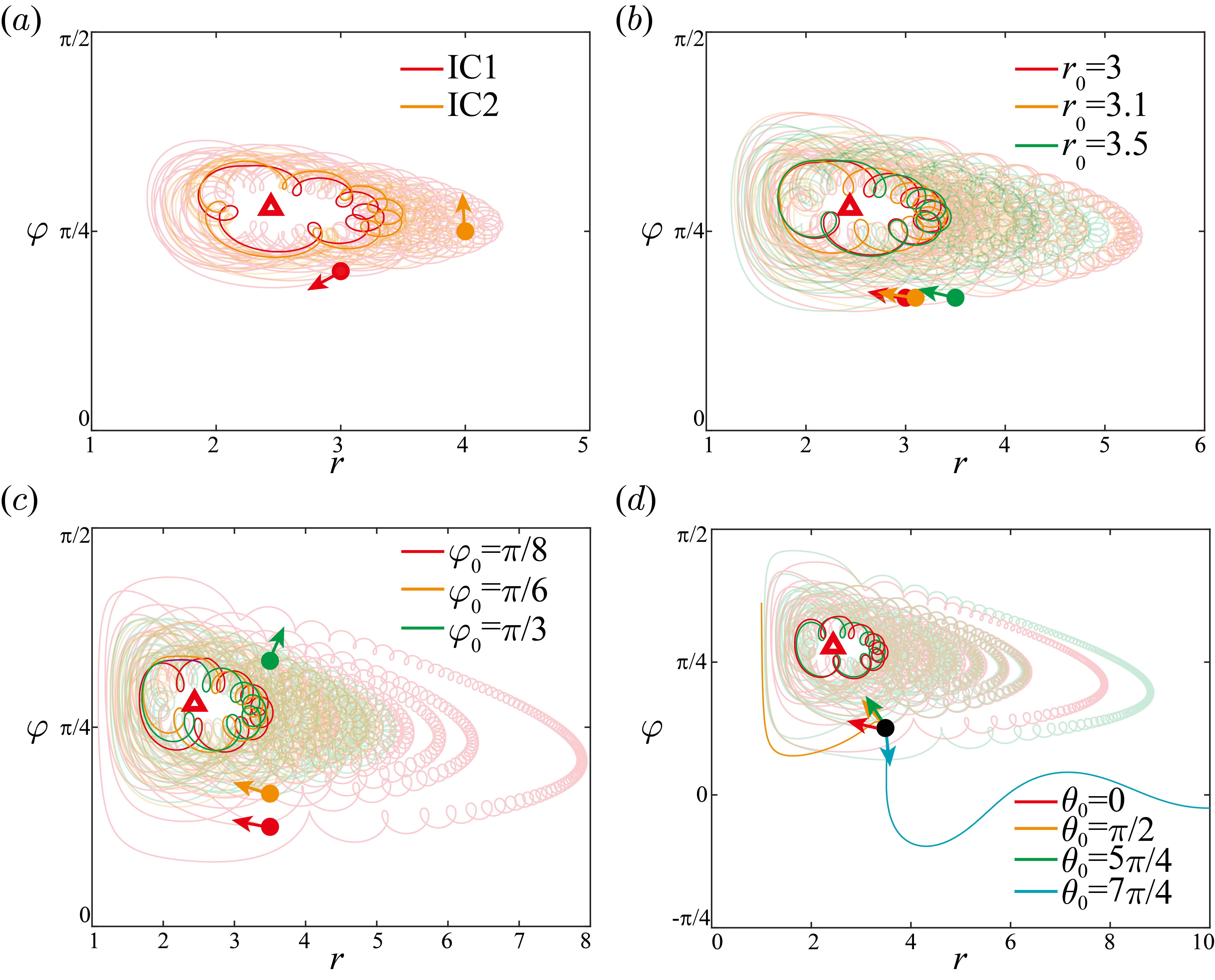}}
            \captionsetup{justification=justified, singlelinecheck=false}
            \caption{(color online) The time evolution results of the radial $r$ and angular position $\varphi$ of active particles with different initial conditions: ($a$) IC1($r_0=3$, $\varphi_0=\pi/5$, $\theta_0=5\pi/4$) and IC2($r_0=4$, $\varphi_0=\pi/4$, $\theta_0=\pi/4$), ($b$) $r_0=3,3.1,3.5, \varphi_0=\pi/6, \theta_0=0$, ($c$) $r_0=3.5, \varphi_0=\pi/8,\pi/6,\pi/3,\theta_0=0$, ($d$) $r_0=3.5, \varphi_0=\pi/8, \theta_0=0,\pi/2,5\pi/4,7\pi/4$. The round dots represent
the initial positions of active particles. Arrows represent their initial motion directions, given by $\left(\dot{r}(r_0,\varphi_0,\theta_0),\dot{\varphi}(r_0,\varphi_0,\theta_0)\right)$. The transparent lines represent the detailed time evolution results, while the solid lines represent their stable orbits. The corresponding parameters are $\alpha=20$, $\gamma=1$.}
             \label{fig11}
           \end{figure}

In this appendix, we include the full time evolution of trajectories for various initial conditions in the $(r, \varphi)$ space, including IC1 and IC2 explicitly mentioned in the text, at $\alpha=20$. As shown in figure \ref{fig11}, the results illustrate the transient dynamics of the active particles, showing how their positions evolve over time before reaching a steady state. We observe that under different initial conditions, the trajectories of active particles undergo varying evolutions. However, the trajectories of captured particles ultimately converge to similar stable bound orbits, which demonstrates the stability of the particle's bounded trajectories. For certain initial conditions in figure \ref{fig11}($d$), particles may either escape the flow field without being captured or collide with the bubble, which also reveals the complexity of particle dynamics within the acoustic streaming flow field.

\bibliographystyle{jfm}
\bibliography{jfm-instructions}

\end{document}